\documentclass[12pt,a4paper]{article}
\usepackage{amsmath,amssymb,eucal}
\usepackage[dvips]{lscape,graphicx}
\newcommand{\bc}{\begin{center}}
\newcommand{\ec}{\end{center}}
\newcommand{\be}{\begin{equation}}
\newcommand{\ee}{\end{equation}}
\newcommand{\ba}{\begin{array}}
\newcommand{\ea}{\end{array}}
\newcommand{\bea}{\begin{eqnarray}}
\newcommand{\eea}{\end{eqnarray}}
\newcommand{\bt}{\begin{tabular}}
\newcommand{\et}{\end{tabular}}

\newcommand{\bsl}{\boldsymbol}

\begin{document}
\ \vspace{20mm}

\begin{center} {\Large\bf Spectroscopy of Baryons
Containing Two Heavy\\[5mm] Quarks  in  Nonperturbative Quark Dynamics}
\end{center}
\vspace{10mm}

\begin{center}I.M.Narodetskii, A.N.Plekhanov,
A.I.Veselov \vspace{2mm}

Institute of Theoretical and Experimental Physics, Moscow
\end{center}\vspace{20mm}

\section*{Abstract}

We have studied the three quark systems in an Effective
Hamiltonian approach in QCD. With only two parameters: the string
tension $\sigma=0.15$ GeV$^2$ and the strong coupling constant
$\alpha_s=0.39$ we obtain a good description of the ground state
light and heavy baryons. The prediction of masses of the doubly
heavy baryons not discovered yet are also given. In particular, a
mass of 3637 MeV for the lightest $ccu$ baryon is found by
employing the hyperspherical formalism to the three quark
confining potential with the string junction.
\vspace{30mm}

The discovery of the $B_c$ meson \cite{PDG} demonstrates that new
sectors of hadron physics are becoming accessible to experiment.
In particular,  the existence of doubly heavy baryons is a natural
consequence of the quark model, and it would be surprising if they
did not exist. Data from the BaBar and Belle collaborations at the
SLAC and KEK B-factories would be good places to look for doubly
charmed baryons. Recently the SELEX, the charm hadroproduction
experiment at Fermilab, reported a narrow state at $3519\pm 1$ MeV
decaying in $\Lambda_c^+K^-\pi^+$, consistent with the weak decay
of the doubly charged baryon $\Xi_{cc}^+$ \cite{Mattson02}. The
SELEX result was recently critically discussed in \cite{KL02}.
Whether or not the state that SELEX reports turns out to be the
first observation of doubly charmed baryons, studying their
properties is important for a full understanding of the strong
interaction between quarks.

Estimations for the masses and spectra of the baryons containing
two or more heavy quarks have been considered by many authors
\cite{BaBar}. The purpose of this letter is to present a
consistent treatment of the masses and wave functions of the
light, heavy and doubly heavy baryons obtained in a simple
approximation within the nonperturbative QCD. In Ref. \cite{Si88}
starting from the QCD Lagrangian and assuming the minimal area for
the asymptotics of the Wilson loop the Hamiltonian of the $3q$
system in the rest frame has been derived. The methodology of the
approach has been reviewed recently \cite{Si02}. By using this
approach and the hypercentral approximation \cite{KNS87} we
calculate the ground state energies and wave functions of the
doubly heavy baryons as three quark systems, with the three-body
confinement force. As a by-product, we also report the masses and
wave functions for light and heavy baryons.


From experimental point of view, a detailed discussion of the
excited QQ'q states is probably premature. Therefore we consider
the ground state baryons without radial and orbital excitations in
which case tensor and spin-orbit forces do not contribute
perturbatively. Then only the spin-spin interaction survives in
the perturbative approximation. The EH has the following form
\cite{Si02}
\begin{equation}
\label{EH} H=\sum\limits_{i=1}^3\left(\frac{m_i^{(0)2}}{2m_i}+
\frac{m_i}{2}\right)+H_0+V,
\end{equation}
where $H_0$ is the non-relativistic kinetic energy operator, $V$
is the sum of the perturbative one-gluon exchange potentials
$V_c$:
\begin{equation}
V_c=-\frac{2}{3}\alpha_s\cdot\sum\limits_{i<j}\frac{1}{r_{ij}},
\end{equation} and the string potential
\begin{equation}
V_{\rm{string}}(\bsl{r}_1,\bsl{r}_2, \bsl{r}_3)=\sigma
l_{\rm{min}},\end{equation}  where $l_{\rm{min}}$ is the sum of
the three distances $|\bsl{r}_i|$ from the string junction point.
In contrast to the standard approach of the constituent quark
model the dynamical masses $m_i$ are no longer free parameters.
They are expressed in terms of the running masses $m^{(0)}_i(Q^2)$
defined at the appropriate hadronic scale of $Q^2$ from the
condition of the minimum of the baryon mass $M_B^{(0)}$ as
function of $m_i$:
\begin{equation} \label{minimum_condition} \frac{\partial M_B^{(0)}(m_i)}{\partial
m_i}=0, ~~~
M_B^{(0)}=\sum\limits_{i=1}^3\left(\frac{m_i^{(0)2}}{2m_i}+
\frac{m_i}{2}\right)+E_0(m_1,m_2,m_3), \end{equation} $E_0$ being
eigenvalue of the operator $H_0+V$.  Technically, this has been
done using the einbein (auxiliary fields) approach, which is
proven to be rather accurate in various calculations for
relativistic systems. Einbeins are treated as $c$ number
variational parameters: the eigenvalues of the EH are minimized
with respect to einbeins to obtain the physical spectrum. Such
procedure
provides the
reasonable accuracy for the meson ground states \cite{mesons_mns}.

The physical mass $M_B$ of a baryon is \cite{Si01_selfenergy}
\begin{equation}\label{self_energy}
M_B=M_B^{(0)}+C,~~~
C=-\frac{2\sigma}{\pi}\sum\limits_i\frac{\eta_i}{m_i},
\end{equation}
where the constant $C$ has the meaning of the quark self energy.
The values of $\eta_i$ are taken from \cite{Si01_selfenergy}. They
are 1, 0.88, 0.234, and 0.052 for $q$, $s$, $c$, and $b$ quarks,
respectively.

The Effective Hamiltonian is solved using the hyperspherical
approach adequate for confining potentials. The baryon wave
function depends on the three-body Jacobi coordinates
\begin{equation}\label{rho}
\bsl{\rho}_{ij}=\sqrt{\frac{\mu_{ij}}{\mu}}(\bsl{r}_i-\bsl{r}_j),~~\bsl{\lambda}_{ij}=\sqrt{\frac{\mu_{ij,k}}{\mu}}
\left(\frac{m_i\bsl{r}_i+m_j\bsl{r}_j}{m_i+m_j}-\bsl{r}_k\right)\end{equation}
($i,j,k$ cyclic), where $\mu_{ij}$ and $\mu_{ij,k}$ are the
appropriate reduced masses
\begin{equation}\mu_{ij}=\frac{m_im_j}{m_i+m_j},~~
\mu_{ij,k}=\frac{(m_i+m_j)m_k}{m_i+m_j+m_k},\end{equation} and
$\mu$ is an arbitrary parameter with the dimension of mass which
drops off in the final expressions. In terms of the Jacobi
coordinates the kinetic energy operator $H_0$ is written as
\begin{equation} \label{H_0_jacobi} H_0= -\frac{1}{2\mu}
\left(\frac{\partial^2}{\partial\bsl{\rho}^2}
+\frac{\partial^2}{\partial\bsl{\lambda}^2}\right)
=-\frac{1}{2\mu}\left( \frac{\partial^2}{\partial
R^2}+\frac{5}{R}\frac{\partial}{\partial R}+
\frac{K^2(\Omega)}{R^2}\right), \end{equation} where $R$ is the
six-dimensional hyper-radius $
R^2=\bsl{\rho}_{ij}^2+\bsl{\lambda}_{ij}^2$, and $K^2(\Omega)$ is
angular momentum operator whose eigen functions (the hyper
spherical harmonics) are $K^2(\Omega)Y_{[K]}=-K(K+4)Y_{[K]}$, with
$K$ being the grand orbital momentum. In terms of $Y_{[K]}$ the
wave function $\psi(\bsl{\rho},\bsl{\lambda})$ can be written in a
symbolical shorthand as \cite{Si66}
\begin{equation}\psi(\bsl{\rho},\bsl{\lambda})=\sum\limits_K\psi_K(R)Y_{[K]}(\Omega).
\end{equation}
In the hyper radial approximation  which we shall use below $K=0$
and $\psi=\psi(R)$. Since $R^2$ is exchange symmetric the baryon
wave function is totally symmetric under exchange.
Introducing the variable $x=\sqrt{\mu} R$ and averaging the
interaction $U=V_c+ V_{\rm{string}}$ over the six-dimensional
sphere $\Omega_6$ one obtains the Schr\"odinger equation for
$u(x)=x^{5/2}\psi(x)$
\begin{equation} \label{shr}
\frac{d^2u(x)}{dx^2}+2\left[E_0+\frac{a}{x}-bx-\frac{15}{8x^2}\right]u(x)=0,
\end{equation}
with the boundary conditions $u(x) \sim {\cal O} (x^{5/2})$ as
$x\to 0$ and the asymptotic   $u(x)\sim
Ai(y)\sim\frac{1}{2}\pi^{-1/2}y^{-1/4}$
$\exp(-\frac{2}{3}y^{3/2})$, $y=(2b)^{1/3}x$,  as $x\to \infty $.
In Eq. (\ref{shr}) $E_0$ is the ground state eigenvalue and
\begin{equation} \label{ab}
\begin{aligned}
a&=\frac{2\alpha_s}{3}\cdot \frac{16}{3\pi}\cdot
\sum\limits_{i<j}\sqrt{\mu_{ij}},\\ b&=\frac{1}{R\sqrt\mu}\int
V_{\rm{string}}(\bsl{r}_1,\bsl{r}_2,\bsl{r}_3)\cdot
\frac{d\Omega_6}{\pi^3},
\end{aligned}
\end{equation}
The potential $V_{\text{string}}(\bsl{r}_1,\bsl{r}_2, \bsl{r}_3)$
has rather complicated structure. In the Y-shape, the string meet
at $120^o$ in order to insure the minimum energy. This shape moves
continuously to a two--legs configuration where the legs meet at
an angle larger than $120^o$.  Let $\varphi_{ijk}$ be the angle
between the line from quark $i$ to quark $j$ and that from quark
$j$ to quark $k$. If $\varphi_{ijk}$ are all smaller than
$120^\circ$, then the equilibrium junction position  coincides
with the so-called Torichelli point of the triangle in which
vertices three quarks are situated. In this case, in terms of the
variables $x$, $\theta=\arctan(\rho_{12}/\lambda_{12})$, and $
\cos\chi=\bsl{\rho_{12}}\cdot\bsl{\lambda_{12}}/\rho_{12}\lambda_{12}$
($0\leqq \theta\leqq \pi/2$, $0\leqq\chi\leqq \pi$) one obtains
\begin{eqnarray}
\l^2_{\text{min}}&=&
x^2\cos^2\theta\\&&\times\left({(m^3_1-m^3_2)\tan^2\theta\over
m_1m_2(m^2_1-m^2_2)}+  \left({m_2-m_1\over
m_2+m_1}\cos\chi+\sqrt3\sin\chi\right){\tan\theta\over
m}+{1\over\mu_{12,3}}\right)\nonumber,
\end{eqnarray}
where $m^2=m_1m_2m_3/(m_1+m_2+m_3).$
 For
the case $m_1=m_2=m_3$ this expression coincides with that derived
in Ref. \cite{FS91}. If $\varphi_{ijk}>120^o$, the lowest energy
configuration
 has the junction at the position
of quark $j$ and $l_{\text{min}}=r_{ij}+r_{jk}$, where
\begin{equation}
r_{12}=\frac{x\sin\theta}{\sqrt{\mu_{12}}},~~
r_{13}=\frac{x\cos\theta}{\sqrt{\mu_{12,3}}}\sqrt{{m^2\over
m_1^2}\tan^2\theta+{2m\over m_1}\tan\theta\cos\chi+1},
\end{equation}
and
\begin{equation}r_{23}=\frac{x\cos\theta}{\sqrt{\mu_{12,3}}}\sqrt{{m^2\over
m_2^2}\tan^2\theta-{2m\over m_2}\tan\theta\cos\chi+1}.
\end{equation}

\begin{figure}
\label{regions}
\begin{center}
\includegraphics
[width=120mm, keepaspectratio=true] {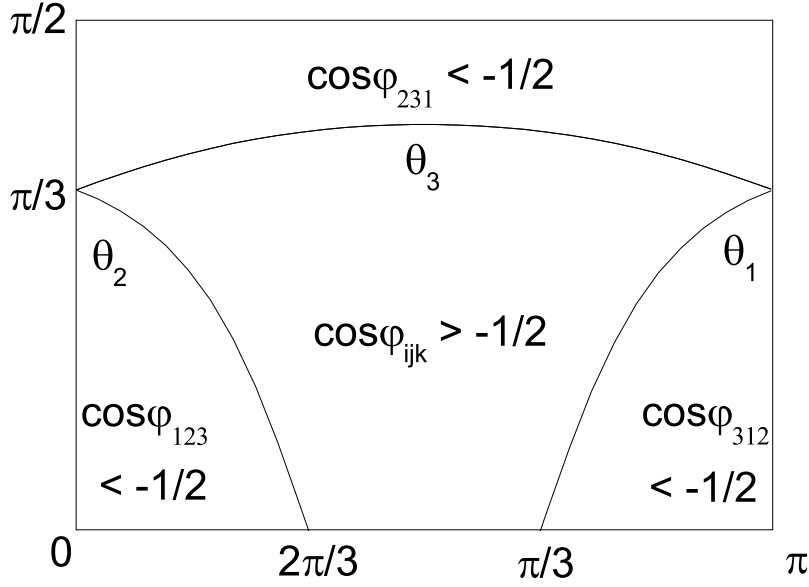}
\end{center}
\caption{The four regions in the $(\theta,\chi)$ plane
corresponding to $\varphi_{ijk}\geq 120^o$ and $\varphi_{ijk} \leq
120^o$ for the case of equal quark masses}
\end{figure}

\noindent The boundaries corresponding to the various regions in
the $(\theta, \chi)$ plane are:
$\theta_{1(2)}(\chi)=\arctan(m_{1(2)}(\mp\cos\chi-\sin\chi/\sqrt3)/m),
$
$\theta_3(\chi)=\arctan(m_2(f(\chi)+\sqrt{f^2(\chi)+4\kappa})/2m),$
$f(\chi)=(1-\kappa)\cos\chi+(1+\kappa)\sin\chi/\sqrt3$,
$\kappa=m_1/m_2$. These boundaries are shown in Fig. 1 for the
case of equal quark masses.

Note that the frequently used approximation \cite{NT01},
\cite{Si02} is to choose the string junction point as coinciding
with the center-of-mass coordinate. In this case
\begin{equation}
\label{cma} V_{\text{string}}=\sigma\sum\limits_{i<j}
\frac{1}{m_k}\cdot\sqrt{\mu\mu_{ij,k}}\cdot
|\bsl{\lambda}_{ij}|,~~b=\sigma\cdot\frac{32}{15\pi}\cdot\sum\limits_{i<j}\frac{\sqrt{\mu_{ij,k}}}{m_k}.
\end{equation}
This approximation that greatly simplifies the calculations
increases the value of $b$ in Eq. (\ref{shr}) by $\sim 5\%$ as
illustrated by the results of Table 1.

We first solve Eq. (\ref{minimum_condition}) for the dynamical
quark masses $m_i$ retaining only the string potential in the
effective Hamiltonian (\ref{EH}). This procedure is consistent
with Ref. \cite{Si02}, but different from that of \cite{NT01}.
Then we add the perturbative Coulomb potential and solve Eq.
(\ref{shr}) to obtain the ground state eigenvalues $E_0$. The
baryon masses $M_B$ are then obtained from solving Eq.
(\ref{self_energy}). \par\vspace{4mm} \noindent {\bf Table 1}.
Illustration of the accuracy of the approximation (\ref{cma}).
Shown are the values $b\sqrt{m_q}/\sigma$ given by Eqs. (\ref{ab})
and (\ref{cma}) where $m_q$ is the light quark mass.

\begin{center}
\begin{tabular}{|c|c|c|} \hline
\ barion & Eq. (\ref{ab}) & Eq. (\ref{cma})
\\ \hline qqq & 1.583 & 1.663
\\ \hline qqs & 1.556 & 1.636
\\ \hline qss & 1.530 & 1.608
\\ \hline qsc & 1.339 & 1.417
\\ \hline qqb & 1.293 & 1.384
\\ \hline qcb & 1.038 & 1.101
\\ \hline qbb & 0.925 & 0.975
\\ \hline
\end{tabular}
\end{center}
\vspace{5mm}

\par
\noindent {\bf Table 2}. For various $3q$ systems in column (1) we
display the dynamical quark masses given by Eq.
(\ref{minimum_condition}), the ground state eigenvalue $E_0$ in
Eq. (\ref{shr}), the baryon masses including the self energy
correction Eq. (\ref{self_energy}) (all in units of GeV) and the
correlation function $\gamma$, Eq. (\ref{gamma}) (in units
GeV$^{3/2}$). \vspace{2mm}

\begin{center}
\begin{tabular}{|c|c|c|c|c|c|c|} \hline
\ baryon & ~~~$m_1$~~~ & ~~~$m_2$~~~ & ~~~$m_3$~~~ & ~~~$E_0$~~~ &
~~~$M_B$~~~& ~~~~$\gamma$~~~~
\\ \hline qqq & 0.362 & 0.362 & 0.362 & 1.392 & ~1.144 & 0.1389
\\ \hline qqs & 0.367 & 0.367 & 0.407 & 1.362 & ~1.242 & 0.1369
\\ \hline qss & 0.371 & 0.411 & 0.411 & 1.335 & ~1.336 & 0.1351
\\ \hline sss & 0.415 & 0.415 & 0.415 & 1.307 & ~1.426 & 0.1333
\\ \hline qqc & 0.406 & 0.406 & 1.470 & 1.142 & ~2.464 & 0.1241
\\ \hline qsc & 0.409 & 0.448 & 1.471 & 1.116 & ~2.542 & 0.1228
\\ \hline ssc & 0.452 & 0.452 & 1.473 & 1.090 & ~2.621 & 0.1214
\\ \hline qqb & 0.425 & 0.425 & 4.825 & 1.054 & ~5.823 & 0.1201
\\ \hline qsb & 0.429 & 0.469 & 4.826 & 1.026 & ~5.903 & 0.1188
\\ \hline ssb & 0.471 & 0.471 & 4.826 & 1.000 & ~5.975 & 0.1177
\\ \hline qcc & 0.444 & 1.494 & 1.494 & 0.876 & ~3.659 & 0.1143
\\ \hline scc & 0.485 & 1.496 & 1.496 & 0.851 & ~3.726 & 0.1134
\\ \hline qcb & 0.465 & 1.512 & 4.836 & 0.753 & ~6.969 & 0.1136
\\ \hline scb & 0.505 & 1.514 & 4.837 & 0.729 & ~7.032 & 0.1128
\\ \hline qbb & 0.488 & 4.847 & 4.847 & 0.567 & 10.214 & 0.1207
\\ \hline sbb & 0.526 & 4.851 & 4.851 & 0.544 & 10.273 & 0.1202
\\ \hline
\end{tabular}
\end{center}
\vspace{5mm}

We use the same parameters as in Ref. \cite{KN00}:
$\sigma=0.15\text{~GeV}^2$ (this value has been confirmed in a
recent lattice study \cite{TMNS02}), $\alpha_s=0.39$,
$m^{(0)}_q=0.009$ GeV, $m^{(0)}_s=0.17$ GeV, $m^{(0)}_c=1.4$ GeV,
and $m^{(0)}_b=4.8$ GeV. In Table 2 for various three-quark
states, we give the quark masses $m_i$, the ground state
eigenvalues $E_0$, and the baryon masses $M_B$. For completeness,
in the last column we report the values of the integral
\begin{equation}\label{gamma} \gamma=\int\frac{u^2(x)}{x^3}dx \end{equation}
in terms of which the quantities
$R_{ijk}=(4\mu_{ij}^{3/2}/\pi^2)\cdot\gamma$ are expressed
determining the probability to find a quark $i$ at the location of
the quark $j$ in a baryon $ijk$. These quantities are of special
importance for the lifetime calculations of heavy hadrons.

Note that there is no good theoretical reason why quark masses
$m_i$ need to be the same in different baryons. Inspection of
Table 2 shows that the masses of the light quarks ($u$, $d$ or
$s$) are increased by $\sim 100$
 MeV when going from the light to heavy baryons. The dynamical
masses of light quarks $m_q\sim\sqrt{\sigma}\sim~400-500$ MeV
qualitatively agree with the results of Ref. \cite{KN00} obtained
from the analysis of the heavy--light ground state mesons.

While studying Table 2 is sufficient to have an appreciation of
the accuracy of our predictions, few comments should be added. We
expect an accuracy of the baryon predictions to be $\sim 5-10\%$
that is partly due to the approximations employed in the
derivations of the Effective Hamiltonian itself \cite{Si02} and
partly due to the error associated with the variational nature of
hyperspherical approximation. From this point of view the overall
agreement with data is quite satisfactory. E.g. we get
$\frac{1}{2}(N+\Delta)_{\text{theory}}~=~1144$ MeV vs
$\frac{1}{2}(N+\Delta)_{\text{exp}}~=~1085$ MeV ( a $5\%$ increase
in $\alpha_s$ would correctly give the $N-\Delta$ center of
gravity), $\frac{1}{4}(\Lambda+\Sigma+2\Sigma^*)~=~1242$ MeV vs
experimental value of 1267 MeV. We also find
$\Xi_{\text{theory}}~=~1336$ MeV (without hyperfine splitting) vs
$\Xi^{1/2}_{\text{exp}}~=~1315$ MeV and $\Xi_{c~
\text{theory}}~=~2542$ Mev vs $\Xi_{c~\text{exp}}~=~$ 2584 MeV. On
the other hand, our study shows some difficulties in reproducing
{\it e.g.} the $\Omega$-hyperon mass.

In Table 3 we compare the spin--averaged masses (computed without
the spin--spin term) of the lowest double heavy baryons to the
predictions of other models \cite{GKLO00}, \cite{EFGM02},
\cite{BDGN94} as well as variational calculations of Ref.
\cite{NT01} for which the center of gravity of non-strange baryons
and hyperons is essential a free parameter. Most of recent
predictions were obtained in a light quark-heavy diquark model
\cite{GKLO00}, \cite{EFGM02}, in which case the spin-averaged
values are $M=\frac{1}{3}(M_{1/2}+M_{3/2})$. Note that the wave
function calculated in the hyperspherical approximation shows the
marginal diquark clustering in the doubly heavy baryons. This is
principally kinematic effect related to the fact that in this
approximation the difference between the various mean values
$\bar{r}_{ij}$ in a baryon is due to the factor
$\sqrt{1/\mu_{ij}}$ which varies between $\sqrt{2/m_i}$ for
$m_i=m_j$ and $\sqrt{1/m_i}$ for $m_i\ll m_j$. In general, in
spite of the completely different physical picture, we find a
reasonable agreement within 100 MeV between different predictions
for the ground state masses of the doubly heavy baryons. Our
prediction for $M_{ccu}$ is 3.66 GeV with the perturbative
hyperfine splitting $\Xi^*_{ccu}-\Xi_{ccu}~\sim~40$ MeV. Note that
the mass of $\Xi_{cc}^+$ is rather sensitive to the value of the
running $c$-quark mass $m_c^{(0)}$ \cite{NT02}.

In conclusion, we have have shown that baryon spectroscopy can be
unified in a single framework of the Effective Hamiltonian which
is consisted with QCD. This picture uses the stringlike picture of
confinement and perturbative one-gluon exchange potential. The
main advantage of this work is demonstration of the fact that it
is possible to describe all the baryons in terms of the only two
parameters inherent to QCD, namely $\sigma$ and $\alpha_s$.
\vspace{5mm}
\par
\noindent {\bf Table 3}. Comparison of our predictions for ground
state masses (in units of GeV) of doubly heavy baryons with other
predictions. \vspace{4mm}

\begin{center}
\begin{tabular}{|c|c|c|c|c|c|} \hline Baryon & This & ~Ref.\cite{NT01}~ &
~Ref.\cite{GKLO00}~ & ~Ref.\cite{EFGM02}~ & ~Ref.\cite{BDGN94}~
\\ & Work & & & &
\\ \hline $\Xi_{cc}$ & ~3.66 & ~3.69 & ~3.57 & ~3.69 & ~3.70
\\ \hline $\Omega_{cc}$ & ~3.73 & ~3.86 & ~3.66 & ~3.84 & ~3.80
\\ \hline $\Xi_{cb}$ & ~6.97 & ~6.96 & ~6.87 & ~6.96 & ~6.99
\\ \hline $\Omega_{cb}$ & ~7.03 & ~7.13 & ~6.96 & ~7.15 & ~7.07
\\ \hline $\Xi_{bb}$ & 10.21 & 10.16 & 10.12 & 10.23 & 10.24
\\ \hline $\Omega_{bb}$ & 10.27 & 10.34 & 10.19 & 10.38 & 10.34
\\ \hline
\end{tabular}
\end{center}
\vspace{5mm}

\noindent This work was supported in part by RFBR grants \#\#
00-02-16363 and 00-15-96786.


\end{document}